\documentclass[conference]{IEEEtran}
\IEEEoverridecommandlockouts
\usepackage{cite}
\usepackage{amsmath,amssymb,amsfonts}
\usepackage{algorithm}
\usepackage[noend]{algpseudocode}
\usepackage{graphicx}
\usepackage{textcomp}
\usepackage{xcolor}
\usepackage[a4paper, total={184mm,239mm}]{geometry}
\def\BibTeX{{\rm B\kern-.05em{\sc i\kern-.025em b}\kern-.08em
    T\kern-.1667em\lower.7ex\hbox{E}\kern-.125emX}}
\begin{document}

\title{A Resource-efficient Spiking Neural Network Accelerator Supporting Emerging Neural Encoding\\
\thanks{This work was supported by the Singapore Government’s Research, Innovation and Enterprise 2020 Plan (Advanced Manufacturing and Engineering domain) under Grant A1687b0033.

Correspondence to Tao Luo (leto.luo@gmail.com)}
}

\author{
    \IEEEauthorblockN{Daniel Gerlinghoff\IEEEauthorrefmark{2},~
                      Zhehui Wang\IEEEauthorrefmark{2},~
                      Xiaozhe Gu\IEEEauthorrefmark{3},~
                      Rick Siow Mong Goh\IEEEauthorrefmark{2},~
                      Tao Luo\IEEEauthorrefmark{2}}

    \IEEEauthorblockA{\IEEEauthorrefmark{2}\textit{Institute of High Performance Computing, Agency for Science, Technology and Research}, Singapore}
    \IEEEauthorblockA{\IEEEauthorrefmark{3}\textit{Future Network of Intelligence Institute, Chinese University of Hong Kong}, Shenzhen, China}
}

\maketitle

\begin{abstract}
Spiking neural networks (SNNs) recently gained momentum due to their low-power multiplication-free computing and the closer resemblance of biological processes in the nervous system of humans. However, SNNs require very long spike trains (up to 1000) to reach an accuracy similar to their artificial neural network (ANN) counterparts for large models, which offsets efficiency and inhibits its application to low-power systems for real-world use cases. To alleviate this problem, emerging neural encoding schemes are proposed to shorten the spike train while maintaining the high accuracy. However, current accelerators for SNN cannot well support the emerging encoding schemes. In this work, we present a novel hardware architecture that can efficiently support SNN with emerging neural encoding. Our implementation features energy and area efficient processing units with increased parallelism and reduced memory accesses. We verified the accelerator on FPGA and achieve 25\% and 90\% improvement over previous work in power consumption and latency, respectively. At the same time, high area efficiency allows us to scale for large neural network models. To the best of our knowledge, this is the first work to deploy the large neural network model VGG on physical FPGA-based neuromorphic hardware.
\end{abstract}

\begin{IEEEkeywords}
Spiking Neural Network, FPGA Accelerator, Neural Encoding
\end{IEEEkeywords}

\section{Introduction}
Spiking neural networks (SNNs) are a promising alternative to conventional artificial neural networks (ANNs) in terms of energy efficiency. While their architectures are similar, the SNN neuron model and its information transmission resemble a biological brain more closely. In SNN, input data are encoded into spike trains, which are sequences of events represented in a binary format. The length of the sequence is equal to the total number of time steps, during which spikes may occur. With the spike train being of binary nature, the neuron model can be implemented in hardware using adders and multiplexers only. The lack of expensive multiplication operations enables the design of low-power devices for machine learning on the edge~\cite{esser2015backpropagation, stromatias2015scalable, luo2018fpga,luo2021nc}. However, for quite a time, SNNs suffered from lower accuracy compared to their ANN counterparts. A lot of research efforts are devoted to enhance the accuracy of SNNs. Recent studies show that the accuracy of large SNNs could be pushed to a level comparable with their ANN counterparts with ANN-to-SNN conversion and very long spike trains~\cite{sengupta2019going}. High spike train lengths, however, deteriorate the latency and energy efficiency achieved by the multiplication-free computation.

Generating spike trains from floating-point numbers is often done using rate-encoding, which relates the spike frequency to the magnitude of the real value. In contrast, radix encoding is an emerging neural encoding scheme that optimizes spike train length while maintaining high accuracy~\cite{wang2021efficient}. As it considers the order of the input spikes, traditional SNN accelerators cannot support radix-encoded models. Therefore, we propose a novel hardware architecture, which supports a wide variety of network structures and, for the first time, shows the scalability necessary to deploy large networks on real FPGA-based neuromorphic hardware. The contribution of this work can be summarized as follows:
\begin{itemize}
    \item A novel hardware architecture with an efficient data flow that minimizes memory accesses and resource requirements for operations on radix-encoded spike trains.
    \item Scalable design that can support large neural network model (VGG) on FPGA-based neuromorphic hardware.
\end{itemize}

\section{Related Work}
Hardware implementations have been used to simulate the behavior of neuromorphic systems, prioritizing biological plausibility over computing efficiency~\cite{jin2008efficient, moore2012bluehive}. On the contrary, there are low-power FPGA implementations, whose event-driven architectures updates the neuron state only at the occurrence of an input spike~\cite{neil2014minitaur, han2020hardware}. While this leads to a high energy efficiency, they are applied to linear layers only.

Recent works show the deployment of convolutional neural networks on SNN hardware. Fang et al.~\cite{fang2020encoding} present an SNN using the spike response model where neurons are interpreted as infinite impulse response filters, implemented on DSP slices in FPGAs. Ju et al.~\cite{ju2020fpga} propose an architecture which reduces data movement of convolution and max-pooling operations by reusing the values of the input feature maps for multiple output values. S2N2~\cite{khodamoradi2021s2n2} uses a single-instruction-multiple-data architecture for classification of time series data, as well as image inputs. The use of binarized weights, however, cannot generalize well.

\section{Hardware Architecture}
\begin{figure}[t]
\centerline{\includegraphics[width=0.48\textwidth]{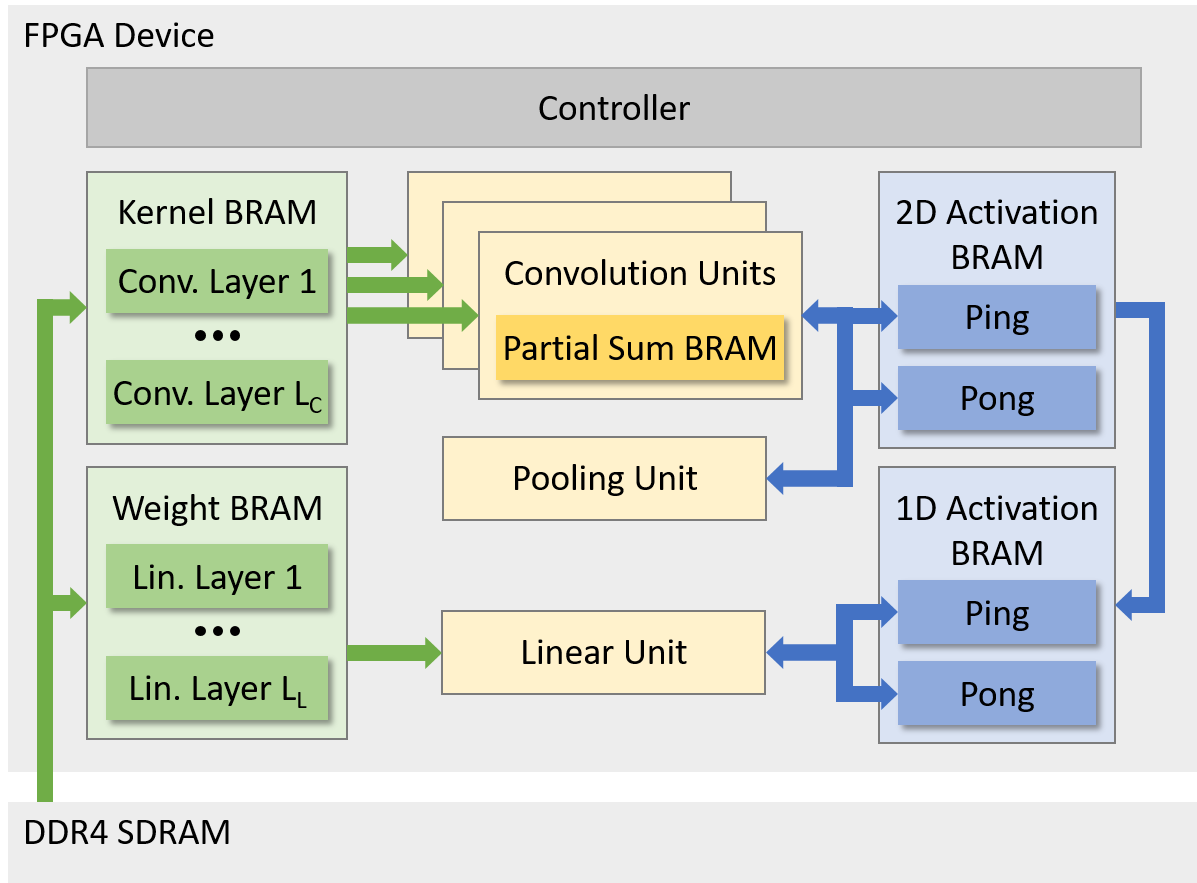}}
\caption{Overview of the accelerator consisting of multiple convolution units, a pooling unit and linear unit (yellow). Model parameters of $L_C$ convolution layers and $L_L$ linear layers are stored in internal block RAM (green) or loaded from external DRAM if internal memory is not sufficient. Activations are stored alternatingly in buffers \textit{ping} or \textit{pong} after each layer (blue). There are separate buffers for 2-dimensional and 1-dimensional activations.}
\label{fig: system overview}
\end{figure}

Our accelerator is able to process a wide variety of user-defined SNN structures. An overview of the architecture is given in Fig.~\ref{fig: system overview}, consisting of three main components: processing units, weight memory, and ping-pong buffers for activations. For each layer, activations and weights are loaded from the respective memory blocks into the processing units, where the neural network operations are applied. The results are written back to the unused activation memory. This execution is managed by a controller.

\subsection{Convolution Unit} \label{sec: convolution unit}
The efficiency of spiking neural networks is rooted in the binarized format of spike trains, obviating the need for multipliers. We propose a novel adder-based architecture for the convolution of radix-encoded spike trains. Adders have a smaller area footprint and power consumption, compared to multipliers and DSP slices. Our data flow allows the reuse of activations and kernels. That minimizes the number of memory accesses and further reduces the power consumption.

\begin{figure}[t]
\centerline{\includegraphics[width=0.44\textwidth]{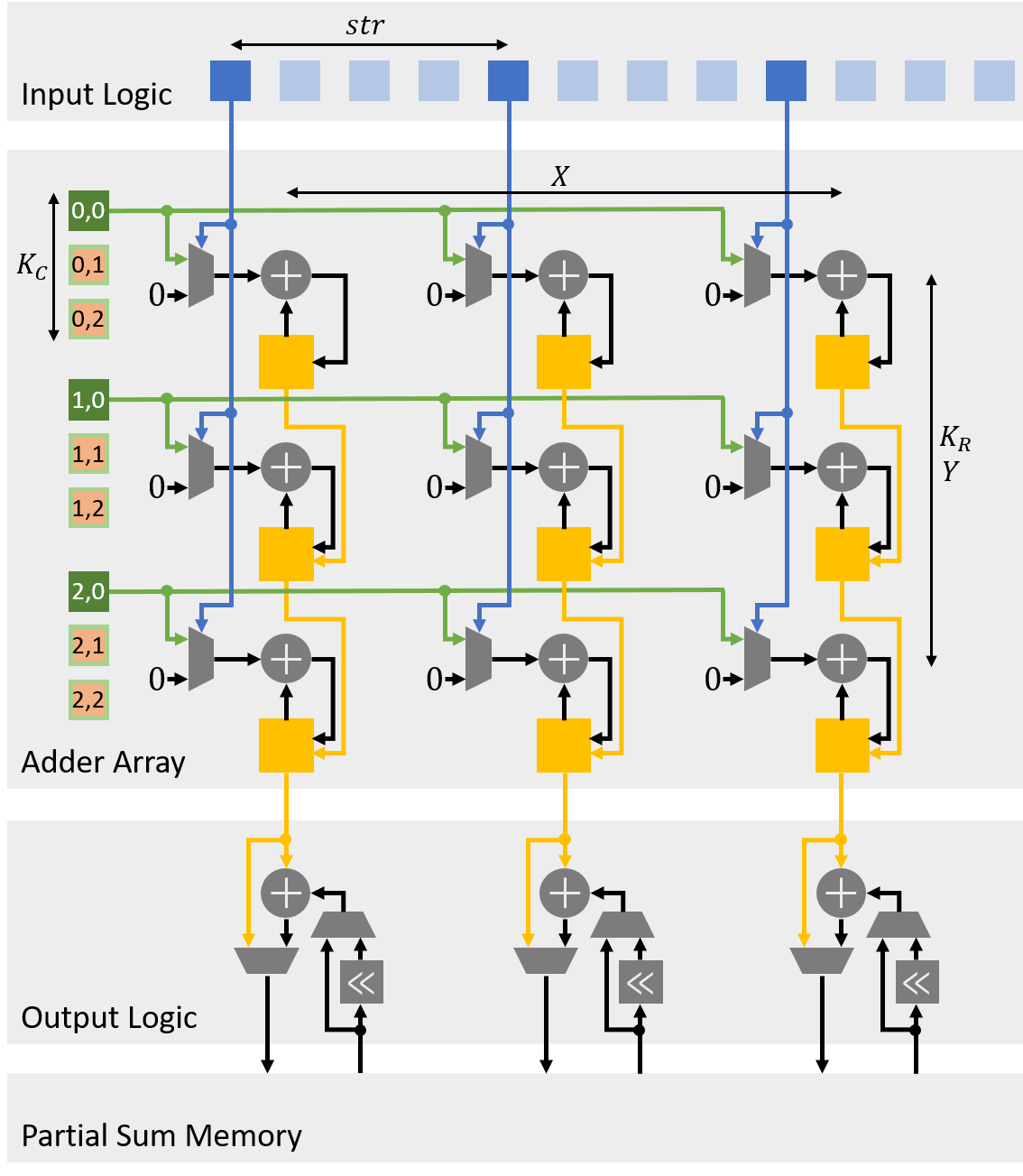}}
\caption{Exemplary components of a convolution unit with kernel rows $K_r=Y=3$ and output size $X=3$. Activations are stored in a shift register in the input logic (blue). Connections to the adder array are established according to stride $str=4$. Every adder row holds the values of the respective kernel row (green). Partial sums are streamed through the adder array from top to bottom (yellow). Input channels and time steps are accumulated by the output logic, where $\ll$ is the left-shift operation.}
\label{fig: convolution unit}
\end{figure}

Our convolution units implement a row-based execution, i.e., all values in a feature map row are processed in parallel. This is reflected in the loop hierarchy by unrolling the activation column loop (see Alg.~\ref{alg: convolution loop}, line~7). The kernel row loop is pipelined, with every row being a separate pipeline stage (line~5). To maximize the utilization of the available resources in the hardware, multiple output channels can share a single convolution unit, if their size permits. A convolution unit is instantiated for one specific kernel size and reused for all layers with the same kernel size. Multiple convolution units with the same kernel size can be created to increase the parallelism with respect to the output channel dimension (line~1).

\begin{algorithm}
\caption{Convolution Loop Hierarchy} \label{alg: convolution loop}
\footnotesize
\begin{algorithmic}[1]
    \ForAll{output channels}\Comment{unroll partially}
    \ForAll{time steps}
    \ForAll{input channels}
    \ForAll{activation rows}
    \ForAll{kernel rows}\Comment{pipeline}
    \ForAll{kernel columns}
    \ForAll{activation columns}\Comment{unroll}
    \State accumulate
    \EndFor
    \State shift input by one column
    \State load next kernel row value
    \EndFor
    \EndFor
    \EndFor
    \State accumulate on previous input channels
    \EndFor
    \State shift sum to the left
    \EndFor
    \State apply ReLU and requantize
    \EndFor
\end{algorithmic}
\end{algorithm}

The unrolling and pipelining of loops is realized in hardware by a two-dimensional array of adders (see Fig.~\ref{fig: convolution unit}). The number of rows $Y$ corresponds to the kernel rows $K_r$, which are computed in parallel. Choosing the number of columns $X$ to be greater or equal than the maximum output channel size can avoid tiling of the feature maps. This does not only help to reduce memory accesses for kernel values, but also reduces the complexity of the control mechanism. The input logic fetches one row of a binary input feature map and writes it into a shift register, which spans the whole length of the row (blue). A connection to the adder array is established for every $str$th value of the input shift register, where $str$ is the kernel stride. For the accumulation along a kernel row, the input array is shifted $K_c$ times, where $K_c$ is the number of kernel columns. This exposes all activation values of the respective kernel windows to the corresponding adders in the adder array.

Coinciding with the shift of the input row, the adder logic loads the new kernel values in the respective kernel row (green). Each adder row $y \in [0, Y-1]$ iterates through the respective kernel row consisting of values $K(y, 0)$ to $K(y, K_c-1)$. This ensures the correct alignment between activation and kernel values during every iteration. The accumulation of kernel values takes place in the adders, if an input spike has occurred. Without the input spike, a multiplexer (gray trapezoid) switches the respective adder input to zero.

After the completion of one kernel row, the partial sums (yellow) are propagated to the next row of adders. The process repeats $K_r$ times. After the partial sum completes the last row, all $K_r*K_c$ kernel values have been applied to each of the $X$ output values.

The result is handed over to the output logic for accumulation with the convolution results of previous input channels and time steps. With radix encoding, a spike at time step $t$ is scaled by a factor of $2^{T-1-t}$, while spikes the following at the following time steps are weighted $2^{T-1-(t+1)}$, with $T$ being the spike train length. To account for this difference, the results computed at $t$ are shifted left by one bit before accumulation with the results at $t+1$ (see Alg.~\ref{alg: convolution loop}, line~12). Partial sums are stored at full integer precision.

\subsection{Pooling and Linear Units}
Pooling units work with the same kind of two-dimensional input data as convolution units and therefore use row-based execution with a similar structure as shown in Fig.~\ref{fig: convolution unit}. Pooling units have a small area footprint, as no kernel values need to be supplied to the adders. Since pooling layers do not involve accumulation over input channels, no dedicated output logic is necessary.

Fully-connected layers use matrix multiplications with a large number of accumulate operations, each requiring an individual weight. To maximize the performance, new weights are fetched at every clock cycle and passed to a row of adders for accumulation over input neurons and time steps. The length of the adder row equals the number of parallel output channels, which is proportional to the available memory bandwidth.

\subsection{Memory Management}
Activations are purely stored in on-chip memory. Two memory blocks exist for two-dimensional and one-dimensional activations, respectively. Convolution and pooling layers load/store their activations using a pair of two-dimensional block RAMs in an alternating manner, also referred to as ping-pong buffer (see Fig.~\ref{fig: system overview}, blue blocks). The width and height of the buffers are determined in a way that minimizes their size while allowing the activations of all relevant layers to fit. Before the computation of the first fully-connected layer, the feature maps are flattened and transferred to the one-dimensional buffers. The the alternation process repeats for all fully-connected layers.

Based on the available on-chip memory resources, there are two available memory options for convolution kernels and weights. If all parameters can be stored in on-chip resources, only on-chip memories are used for every convolution and fully-connected layer. Else, parameters are fetched from off-chip DRAM before the computation of each layer.

\section{Experiments}
\begin{table}[t]
    \begin{minipage}[t]{.39\linewidth}
        \renewcommand{\arraystretch}{1.1}
        \centering
        \setlength{\tabcolsep}{5pt}
        \caption{Accuracy \& Latency\\versus Time Steps}
        \begin{tabular}{c|c c}
            \hline
            \textbf{Time} & \textbf{Acc.} & \textbf{Lat.} \\
            \textbf{Steps} & \textbf{[\%]} & \textbf{[\textmu s]} \\
            \hline
            3 & 98.57 & 648  \\
            4 & 99.09 & 856  \\
            5 & 99.21 & 1063 \\
            6 & 99.26 & 1271 \\
            \hline
        \end{tabular}
        \label{tab: accuracy latency}
    \end{minipage}
    \hfill
    \begin{minipage}[t]{.59\linewidth}
        \renewcommand{\arraystretch}{1.1}
        \centering
        \setlength{\tabcolsep}{5pt}
        \caption{Latency, Power \& Resources\\versus Convolution Units}
        \begin{tabular}{c|c c c c}
            \hline
            \textbf{Conv.} & \textbf{Lat.} & \textbf{Pow.} & \multicolumn{2}{c}{\textbf{Resources}} \\
            \textbf{Units} & \textbf{[\textmu s]} & \textbf{[W]} & \textbf{LUTs} & \textbf{FF} \\
            \hline
            1 & 1063 & 3.07 & 11k & 10k \\
            2 & 648  & 3.09 & 15k & 14k \\
            4 & 450  & 3.17 & 24k & 23k \\
            8 & 370  & 3.28 & 42k & 39k \\
            \hline
        \end{tabular}
        \label{tab: convolution units}
    \end{minipage}
\end{table}

\begin{table*}[t]
    \begin{center}
    \caption{Efficiency and Performance of SNN Hardware Accelerators}
    \renewcommand{\arraystretch}{1.1}
    \begin{tabular}{l l l c|r r r r c}
        \hline
        \textbf{Platform} & \textbf{Dataset} & \textbf{Network} & \textbf{Accuracy} & \textbf{Frequency} & \textbf{Latency} & \textbf{Throughput} & \textbf{Power} & \textbf{LUTs / FF} \\
        {} & {} & {} & \textbf{[\%]} & \textbf{[MHz]} & \textbf{[\textmu s]} & \textbf{[fps]} & \textbf{[W]} & {} \\
        \hline
        Ju et al. \cite{ju2020fpga}         & MNIST     & CNN\textsuperscript{1} & 98.9 & 150  & 6110  & 164  & 4.6  & 107k / 67k  \\
        Fang et al. \cite{fang2020encoding} & MNIST     & CNN\textsuperscript{2} & 99.2 & 125  & 7530  & 2124 & 4.5  & 156k / 233k \\
        \textbf{This work}                  & MNIST     & CNN\textsuperscript{2} & 99.3 & \textbf{200}  & \textbf{409}   & \textbf{2445} & \textbf{3.6}  & \textbf{41K / 36K}   \\
        \hline
        \textbf{This work}                  & MNIST     & LeNet-5                & 99.1 & 200  & 294   & 3380 & 3.4  & 27k / 24k   \\
        \textbf{This work}                  & CIFAR-100 & VGG-11                 & 60.1 & 115  & 210k  & 4.7  & 4.9  & 88k / 84k   \\
        \hline
    \end{tabular}
    \label{tab: hardware performace}
    \end{center}
    \hspace*{16.5mm}\textsuperscript{1} \footnotesize{28×28 -- 64C5 -- 2P -- 64C5 -- 2P -- 128 -- 10},
    \textsuperscript{2} \footnotesize{28x28 -- 32C3 -- P2 -- 32C3 -- P2 -- 256 -- 10}
    \vspace{-10pt}
\end{table*}

\subsection{Experiment Setup}
To verify the accelerator's functionality and evaluate its performance, we deployed it on a Xilinx Virtex UltraScale+ XCVU13P FPGA. As there is no need for multipliers, the arithmetic was implemented in the carry logic and lookup tables of the FPGA fabric, opposed to DSP slices. SystemVerilog was used to describe the hardware, which was directly synthesized and implemented by the Vivado toolchain using default settings.

We evaluate our design with the MNIST dataset on \mbox{LeNet-5}, which has the following architecture: 32x32x1 -- 6C5 -- P2 -- 16C5 -- P2 -- 120C5 -- 120 -- 84 -- 10. SNN models are obtained by the ANN-to-SNN conversion method, which trains an equivalent ANN and transfers the network parameters to an SNN model~\cite{gerlinghoff2021e3ne}. The resolution of the network parameters is set to 3 bits. We set $(X, Y) = (30, 5)$ for convolution units and $(X, Y) = (14, 2)$ for pooling units, according to the network configuration. The two experiments in Sec.~\ref{sec: spike train length} and~\ref{sec: number convolution units} run at a clock frequency of 100 MHz.

In Section~\ref{sec: performance scalability}, we compare our system performance with previous works by deploying the convolutional SNN model of Fang et al.~\cite{fang2020encoding} on our accelerator. Additionally, LeNet and VGG-11 are used to evaluate the scalability for different model sizes. For the small LeNet, four time steps are required to reach acceptable accuracy. We chose four convolution units as they yielded one of the best latency-power-resource ratio in the experiment in Section~\ref{sec: number convolution units}. The accelerator is run at a clock speed of 200 MHz. VGG-11 is a convolutional neural network (CNN), which has 28.5~million parameters and consists of 11 convolution, pooling, or fully-connected layers. It is used for the classification of 100 different objects in the \mbox{CIFAR-100} dataset. Due to the increased complexity, six time steps are needed to maintain the accuracy while eight convolution units are used to reduce the processing time. The accelerator is clocked at 115 MHz.

\subsection{Length of Spike Train} \label{sec: spike train length}
A common trade-off for SNN hardware is between spike train length and accuracy. As the encoding error decreases with an increasing number of time steps, we expect the classification accuracy to improve. Since that affects the execution time, we additionally observe the latency with two parallel convolution units.

In Table~\ref{tab: accuracy latency}, the benefits of radix encoding become clear, as with six time steps, the state-of-the-art accuracy can be reached. Fang et al.~\cite{fang2020encoding} required approximately ten time steps to reach the same accuracy. Hence, a potential efficiency improvement of around 40\% can be achieved by the neural encoding scheme alone. Using more than six time steps lead to no significant improvement of the accuracy. The latency scales linearly with the length of the spike train since almost all computations are replicated for each time step.

\subsection{Number of Convolution Units} \label{sec: number convolution units}
The execution time can be reduced by duplicating convolution units at the expense of silicon area and power consumption. The classification result is unaffected by the number of convolution units as the operations are identical. We use a spike train length of $T = 3$.

Table~\ref{tab: convolution units} shows an improvement in latency with an increasing number of convolution units. However, doubling the convolution units does not lead to reduction of latency by 50\%. This is due to a growing portion of the execution time being used to access activations and kernels. The second reason for the convergence of latency is that the pooling and linear units are not duplicated, since they are already largely constrained by memory bandwidth. On the contrary, hardware resources scale almost linear with the number of convolution units.

\subsection{Hardware Performance and Scalability} \label{sec: performance scalability}
The results and a comparison with previous state-of-the-art implementations are given in Table~\ref{tab: hardware performace}. To allow for a fair comparison, we deployed the convolutional SNN of Fang et al.~\cite{fang2020encoding} on our accelerator. In their paper, they proposed a framework for the generation of SNN hardware which is based on a high-level synthesis flow. This comes at the expense of hardware resources, using up almost 4$\times$ of lookup tables (LUTs) and 6$\times$ of flip-flops (FF) compared to our work. Despite reaching a lower accuracy, they exceed our latency 18-fold and the power consumption by 25\%. Ju et al.~\cite{ju2020fpga} implemented their SNN engine in the programmable logic of a Xilinx Zynq FPGA. We improved the throughput by 15$\times$, while only consuming 75\% of power and less than half of their hardware resources.

On LeNet, we achieve a high classification accuracy of 99.09\% at only 294 \textmu s of latency and a throughput of over 3300 frames per second. Thanks to the emerging neural encoding, our design has high scalability in terms of power, latency and resources, which also makes us the first work that deploys the large SNN model VGG-11 on physical neuromorphic hardware. Due to its network size, accesses to the external DRAM are necessary, which contributes to an increase in power and hardware utilization. Furthermore, a total of 4.5 MB of internal BRAM are required to store the intermediate feature maps. Nevertheless, a throughput of more than four images per second can be achieved.

\section{Conclusion}
In our work, we proposed a novel hardware architecture for convolutional spiking neural networks, which use radix encoding to shorten spike trains while maintaining state-of-the-art accuracy. We adopted a row-based architecture that heavily reduces the number of memory accesses to load kernels and activations. The parallel execution within and between processing units enables low resource usage and high performance. The experiments and comparison demonstrated its versatility and advantages over previous convolutional SNN accelerators in terms of all metrics. Our design achieved high scalability, which allows us to deploy the large VGG model on the physical FPGA-based neuromorphic hardware.

\bibliographystyle{IEEEtran}
\bibliography{date}

\end{document}